\documentclass[%
 reprint,
 amsmath,amssymb,
 aps,
floatfix,
]{revtex4-1}

\usepackage{graphicx}
\usepackage{dcolumn}
\usepackage{bm}
\usepackage[dvips]{color}

\begin{document}


\title{Charge asymmetry from CP-violating fermion scattering off bubble walls during the electroweak phase transition.}

\author{Alejandro Ayala$^{1,2}$, L. A. Hern\'andez$^1$, Jordi Salinas$^1$}
\affiliation{$^1$Instituto de Ciencias
  Nucleares, Universidad Nacional Aut\'onoma de M\'exico, Apartado
  Postal 70-543, M\'exico Distrito Federal 04510,
  Mexico.\\
  $^2$Centre for Theoretical and Mathematical Physics, and Department of Physics,
  University of Cape Town, Rondebosch 7700, South Africa.}
  
\begin{abstract}

We compute the net electric current generated during a first order electroweak phase transition when fermions transit from the false to the true vacuum. This current is generated by the CP-violating fermion interaction with the Higgs during the phase transition and is quantified in terms of a CP-violating phase in the bubble wall separating the symmetric from the symmetry-broken phases. We comment on the seed magnetic field that this current is able to generate and it is possible implications for magnetogenesis in the early universe during the electroweak phase transition.

\begin{description}
\item[PACS numbers]
98.80.Cq, 11.30.Er, 95.85.Sz
\item[Keywords]
Electroweak phase transition, CP-violation, Charge asymmetry, Magnetogenesis 
\end{description}
\end{abstract}

\pacs{Valid PACS appear here}
\maketitle


\section{\label{I}Introduction}

An outstanding problem in modern cosmology is to understand the origin of the observed large scale magnetic fields~\cite{observe}. It has been suggested that these fields could be a relic from the universe's evolution during its early stages. Several scenarios are plausible and among them, phase transitions are one of the venues that is under active consideration. One of such transitions is the electroweak phase transition (EWPT) that took place around temperatures of order 100 GeV. It is by now understood that the EWPT is weakly first order~\cite{Carrington}, and can become stronger if additional ingredients are considered, such as the presence of magnetic fields themselves during the electroweak epoch~\cite{additional}.

One of the first explored mechanisms to generate magnetic fields during a first order phase transition is the so called battery effect~\cite{Hogan}. The generation of electromagnetic fields from gradients of the Higgs field during the EWPT~\cite{Vachaspati} has been also studied in detail. Spontaneous magnetization of Bose-Einstein condensed $W$-bosons in the broken phase after the EWPT has also been explored~\cite{Dolgov}. For reviews see Refs.~\cite{revs}.

A first order EWPT happens through the nucleation of the true vacuum bubbles within the false phase. These bubbles expand and fill up the entire space until the phase transition is completed. Bubble collisions can provide turbulence that can in turn amplify any seed magnetic field by a Dynamo effect~\cite{Stevens}. During the phase transition, particles acquire their masses when passing from the false to the true vacuum through the bubble walls. This transit can be regarded as a quantum scattering process whereby particles in general, and fermions in particular, have a probability of being either transmitted or reflected. 

The problem of fermion scattering off bubble walls has been the subject of several studies in the context of the EWPT. In the absence of any source of asymmetric scattering the solution has been found in full detail for the thin wall regime in Ref.~\cite{AJMV}, by solving the Dirac equation using for the fermion mass a term proportional to the Higgs field profile between degenerate vacua. The CP-violating scenario has been discussed in Ref.~\cite{Funakubo} including CP-violation as part of the bubble wall and finding the solution to the Dirac equation as a perturbation on top of CP-conserving solutions. More recently, the exact solution of the Dirac equation for the case when the Higgs field profile is of the kink type and contains a CP-violating phase, was found in Ref.~\cite{Prokopec}. The P-odd problem in the presence of an external hypermagnetic field, where the asymmetric scattering is provided by the different hypercarge couplings of left- and right-fermion modes, has also been solved in Refs.~\cite{Ayala2}. The emphasis of these works has been the search for extra sources of baryogenesis.

When the scattering of conjugate (in a given charge) fermion modes is asymmetric, the process can give rise to currents whose fate depends on the dynamical properties of the plasma. For instance, it has been found that in the presence of a magnetic field, an imbalance between the number density of left- and right-handed fermions leads to an induced electric current~\cite{Boyarsky} that in turn serves a seed for the development of a magnetic field whose helicity can be inherited from that of the fermions~\cite{helicity}. In this context, it is interesting to explore and quantify the current that CP-violating interactions with the bubble walls induce during the EWPT, as a possible seed for magnetogenesis. In this work we undertake such exploration. We compute the transmission and reflection coefficients for fermion modes incident from the false vacuum. From these coefficients we compute the strength of the generated electric current that is quantified in terms of a CP-violating phase introduced as a property of the bubble wall interacting with fermions. Since the emphasis is on the generated current, we work in the infinite thin wall case to simplify the analysis.  

The work is organized as follows: In section II we write the Dirac equation for the left- and right-handed chirality modes propagating through a zero width wall during the EWPT. CP-violation is introduced in terms of a complex mass term. In section III, we find the solutions in the broken and symmetric phases and discuss their properties. In section IV, we compute the transmission and reflection coefficients and the asymmetry caused by the complex mass term in the Dirac equation. Finally in section V, we summarize and conclude providing possible implications of this asymmetry for primordial magnetogenesis during the EWPT.  

\section{\label{II}Dirac equation with a complex mass}

A first order phase transition happens through bubble nucleation. The interface separating the two 
phases is called the wall. During the EWPT, the properties of the wall depend on those of the effective electroweak effective potential. In the thin wall regime and when the phase transition is considered for degenerate energy 
densities between the false and true vacua, it is possible to find a one-dimensional solution for the Higgs field $\varphi$ called the \textit{kink}. This is given by
\begin{equation}
 \varphi(z)\sim 1+\tanh\left(\frac{z}{\lambda}\right),
 \label{kink}
\end{equation}
where $z$ is the direction along the phase change and $\lambda$ is the wall's width. We consider the limit where the wall has zero width ($\lambda =0$). Therefore the problem of fermion reflection and transmission through the wall can be cast in terms of solving the Dirac equation with a position dependent fermion mass $m(z)$, proportional to the Higgs field. Since the wall has zero width, the kink solution becomes a step function $\Theta (z)$. To allow for a CP-violating interaction between the Higgs and fermions, $m(z)$ is allowed to be complex. Therefore, the expression for the particle's mass becomes explicitly
\begin{equation}
 m(z)=m_0 e^{i\phi} \Theta (z),
 \label{complexmass}
\end{equation}
where $\phi$ is a phase.  CP-violation implies that left- and right-handed chirality modes of a fermion spinor $\Psi$, namelly,
\begin{align}
 \psi_R&= \frac{1}{2}(1+\gamma_5)\Psi, \nonumber \\
 \psi_L&= \frac{1}{2}(1-\gamma_5)\Psi,
 \label{spinorRL}
\end{align}
couple to the Higgs field with $m(z)$ and $m^*(z)$, respectively. The Dirac equation is written as
\begin{equation}
 \Big \{ i\gamma^\mu \partial_\mu -m^*(z)\frac{1}{2}(1-\gamma_5)-m(z)\frac{1}{2}(1+\gamma_5) \Big \}\Psi=0.
 \label{diraceq}
\end{equation}
Hereafter, we work with the chiral representation of the gamma matrices, where
\begin{equation}
 \gamma^0= \begin{pmatrix} 0 & -I \\ -I & 0 \end{pmatrix}, 
 \gamma_j=\begin{pmatrix} 0 & \sigma_j \\ -\sigma_j & 0 \end{pmatrix},
 \gamma_5=\begin{pmatrix} I & 0 \\ 0 & -I \end{pmatrix}.
 \label{gammamatrices}
\end{equation}
We now proceed to find the solution of Eq.~(\ref{diraceq}).

\section{\label{III}Solving the Dirac equation}

A general solution of Eq.~(\ref{diraceq}) is of the form 
\begin{equation}
 \Psi= \Big \{ i\gamma^\mu \partial_\mu +m^*(z)\frac{1}{2}(1-\gamma_5)+m(z)\frac{1}{2}(1+\gamma_5) \Big \}\varPhi.
 \label{soldiraceq}
\end{equation}
Substituting Eq.~(\ref{soldiraceq}) into Eq.~(\ref{diraceq}), we obtain four Klein-Gordon-like equations
\begin{align}
 &\Big \{ -\partial^2 -i\gamma^3m_0e^{-i\phi}\delta(z)\frac{1}{2}(1-\gamma_5) \nonumber \\
 &-i\gamma^3m_0e^{i\phi}\delta(z)\frac{1}{2}(1+\gamma_5)-m_0^2\Theta(z) \Big\}\varPhi=0.
 \label{KGlike}
\end{align}
Using separation of variables, the solution of Eq.~(\ref{KGlike}) can be written as
\begin{equation}
 \varPhi(\bar{x},t)=\xi(x,y)\Phi(z)\eta(t).
\end{equation}
We look for stationary solution, $\eta(t)=e^{-iEt}$, describing the motion of fermions perpendicular to the wall, \textit{i.e.}, along the $\hat{z}$ axis. Thus, Eq.~(\ref{KGlike}) becomes
\begin{align}
 &\Bigg \{ E^2-m_0^2\Theta(z)+\frac{d^2}{dz^2}\nonumber \\
 &-im_0\delta(z)\gamma^3\Bigg[ e^{-i\phi}\frac{1}{2}(1-\gamma_5)+e^{i\phi}\frac{1}{2}(1+\gamma_5) \Bigg] \Bigg \}\Phi(z)=0.
 \label{diraceqz}
\end{align}
Now, we expand $\Phi(z)$ in terms of the eigenspinors $u^s_\pm$ of $\gamma^3$ 
\begin{equation}
 u_\pm^1=\begin{pmatrix}  1 \\ 0 \\ \pm i \\ 0  \end{pmatrix},
 \\\\\ u_\pm^2=\begin{pmatrix}  0 \\ 1 \\ 0 \\ \mp i  \end{pmatrix}, 
 \label{spinorsz}
\end{equation}
where $s=1,2$ are spin indices and $\pm$ label positive and negative energies, respectively. Using Eq.~(\ref{spinorsz}), we can write $\Phi(z)$ as
\begin{equation}
 \Phi(z)=\varPhi_+^1(z)u_+^1+\varPhi_-^1(z)u_-^1+\varPhi_+^2(z)u_+^2+\varPhi_-^2(z)u_-^2,
 \label{decomposition}
\end{equation}
where $u_\pm^1$ and $u_\pm^2$ satisfy the following relations
\begin{align}
 \gamma^3u_\pm^{1,2}&=\pm iu_\pm^{1,2}, \nonumber \\
 \gamma^0u_\pm^1&=\mp iu_\mp^1, \nonumber \\
 \gamma^0u_\pm^2&=\pm iu_\mp^1, \nonumber \\
 \gamma^5u_\pm^{1,2}&=u_\mp^{1,2}.
\end{align}
Substituting Eq.~(\ref{decomposition}) into Eq.~(\ref{diraceqz}) and taking into account that the eigenspinors are mutually orthogonal, we obtain four differential  equations, one for each of the four combinations of $u_\pm^{1,2}$,
\begin{align}
 &\left[E^2-m_0^2\Theta(z)+\frac{d^2}{dz^2}+m_0\delta(z)\cos\phi\right]\varPhi_+^1(z) \nonumber \\
 &=-im_0\delta(z)\sin\phi \ \varPhi_-^1(z), \nonumber \\
 &\left[E^2-m_0^2\Theta(z)+\frac{d^2}{dz^2}-m_0\delta(z)\cos\phi\right]\varPhi_-^1(z) \nonumber \\
 &=im_0\delta(z)\sin\phi \ \varPhi_+^1(z), \nonumber \\
 &\left[E^2-m_0^2\Theta(z)+\frac{d^2}{dz^2}+m_0\delta(z)\cos\phi\right]\varPhi_+^2(z) \nonumber \\
 &=-im_0\delta(z)\sin\phi \ \varPhi_-^2(z), \nonumber \\
 &\left[E^2-m_0^2\Theta(z)+\frac{d^2}{dz^2}-m_0\delta(z)\cos\phi\right]\varPhi_-^2(z) \nonumber \\
 &=im_0\delta(z)\sin\phi \ \varPhi_+^2(z).
 \label{foureq}
\end{align}
To solve the set of Eqs.~(\ref{foureq}), we notice that it is enough to work with only two of the four equations, namely 
$\varPhi_\pm^1$, given that the other two equations are identical. We take
\begin{align}
 &\left[E^2-m_0^2\Theta(z)+\frac{d^2}{dz^2}+m_0\delta(z)\cos\phi\right]\varPhi_+^1(z) \nonumber \\
 &=-im_0\delta(z)\sin\phi \ \varPhi_-^1(z), \nonumber \\
 &\left[E^2-m_0^2\Theta(z)+\frac{d^2}{dz^2}-m_0\delta(z)\cos\phi\right]\varPhi_-^1(z) \nonumber \\
 &=im_0\delta(z)\sin\phi \ \varPhi_+^1(z).
 \label{eqz1}
\end{align}
To find the explicit solution for $\varPhi(z)$, we divide the space direction perpendicular to the wall in three regions, $z<0$, $z=0$ and $z>0$. The solutions are found in each region and then matched, imposing continuity of the solution and accounting for the discontinuity of the first derivative at $z=0$.

When $z<0$, the mass term is equal to zero and thus, the differential equations decouple and can be written as
\begin{align}
 \left[E^2+\frac{d^2}{dz^2}\right]\varPhi_+^1(z)&=0, \nonumber \\
 \left[E^2+\frac{d^2}{dz^2}\right]\varPhi_-^1(z)&=0,
 \label{eqzuncoupled}
\end{align}
whose solutions are
\begin{equation}
 \varPhi_\pm^1(z)=A_\pm e^{iEz}+B_\pm e^{-iEz}.
 \label{solmenos}
\end{equation}
For $z>0$, once again the equations decouple and they are explicitly
\begin{align}
 \left[ E^2-m_0^2+\frac{d^2}{dz^2}\right]\varPhi_+^1(z)&=0, \nonumber \\
 \left[ E^2-m_0^2+\frac{d^2}{dz^2}\right]\varPhi_-^1(z)&=0,
 \label{eqzuncoupledmass}
\end{align}
whose solutions are
\begin{equation}
 \varPhi_\pm^1(z)=F_\pm e^{i\sqrt{E^2-m_0^2}z}+G_\pm e^{-i\sqrt{E^2-m_0^2}z}.
 \label{solmas}
\end{equation}
From the solutions at $z\neq0$, we now impose continuity at $z=0$ which results in the conditions
\begin{equation}
 A_\pm + B_\pm = F_\pm + G_\pm.
 \label{continuity}
\end{equation}
To compute the discontinuity, we first find the derivative of Eqs.~(\ref{solmenos}) and~(\ref{solmas}). These are
\begin{align}
 (z<0) \ \ \ \frac{d}{dz} \varPhi_\pm^1(z)&=iE \big (A_\pm e^{iEz}- B_\pm e^{-iEz} \big ), \nonumber \\
 (z>0) \ \ \ \frac{d}{dz} \varPhi_\pm^1(z)&=i\sqrt{E^2-m_0^2} \big (F_\pm e^{i\sqrt{E^2-m_0^2}z} \nonumber \\
 &-G_\pm e^{-i\sqrt{E^2-m_0^2}z} \big ),
 \label{firstder}
\end{align}
respectively. Evaluating Eq.~(\ref{firstder}) at $z=0$, we get
\begin{align}
 (z<0) \ \ \ \frac{d}{dz} \varPhi_\pm^1(0)&=iE(A_\pm -B_\pm), \nonumber \\
 (z>0) \ \ \ \frac{d}{dz} \varPhi_\pm^1(0)&=i\sqrt{E^2-m_0^2}(F_\pm-G_\pm).
 \label{firstder0}
\end{align}
The proper way to compute the discontinuity at $z=0$ is to integrate Eq.~(\ref{eqz1}) from $0-\epsilon$ to $0+\epsilon$ and then take the limit $\epsilon\rightarrow 0$. In this way we obtain
\begin{align}
 \Delta \left[\frac{d}{dz} \varPhi_+^1(z)\right]&=-m_o \cos \phi \ \varPhi_+^1(0)-im_o \sin \phi \ \varPhi_-^1(0), \nonumber \\
 \Delta\left[\frac{d}{dz} \varPhi_-^1(z)\right]&=m_o \cos \phi \ \varPhi_-^1(0)+im_o \sin \phi \ \varPhi_+^1(0).
 \label{difference}
\end{align}
By using Eqs.~(\ref{solmenos}) and~(\ref{firstder0}) into Eq.~(\ref{difference}), we get the result
\begin{align}
 &A_+(-m_o\cos \phi +iE)+B_+(-m_o\cos \phi-iE) \nonumber \\
 &=i\sqrt{E^2-m_0^2}(F_+-G_+)+im_0\sin \phi (A_-+B_-), \nonumber \\
 &A_-(m_0\cos \phi+iE)+B_-(m_0\cos \phi-iE) \nonumber \\
 &=i\sqrt{E^2-m_0^2}(F_-G_-)-im_0\sin \phi (A_++B_+).
 \label{discontinuity}
\end{align}
Notice that we are left with two sets each with two differential equations, Eqs.~(\ref{continuity}) and~(\ref{discontinuity}), and eight unknown coefficients, $A_\pm$, $B_\pm$, $F_\pm$ and $G_\pm$. However, if we assume that the incident wave comes from $z \rightarrow -\infty$, then after the wave crosses the wall, there is only one resulting wave traveling in the same direction, hence $G_\pm=0$ and the system of equations is reduced to
\begin{align}
 &A_+ + B_+ = F_+, \nonumber \\
 &A_- + B_- = F_-, \nonumber \\
 &A_+(-m_o\cos \phi +iE)+B_+(-m_o\cos \phi-iE) \nonumber \\
 &=i\sqrt{E^2-m_0^2}(F_+)+im_0\sin \phi (A_-+B_-), \nonumber \\
 &A_-(m_0\cos \phi+iE)+B_-(m_0\cos\phi-iE) \nonumber \\ &=i\sqrt{E^2-m_0^2}(F_-)-im_0\sin \phi (A_++B_+).
\end{align}
Since the incident wave comes from $z\rightarrow -\infty$, we express the solution in terms of $A_\pm$, such that the other coefficients are written as
\begin{align}
 B_-&=A_+\left[ \frac{m_0\sin \phi}{E+\sqrt{E^2-m_0^2}}\right]-A_- \left[ \frac{im_0\cos \phi}{E+\sqrt{E^2-m_0^2}} \right], \nonumber \\
 B_+&=-A_+ \left[ \frac{im_0\cos \phi}{E+\sqrt{E^2-m_0^2}} \right]-A_-\left[ \frac{m_0\sin \phi}{E+\sqrt{E^2-m_0^2}} \right],
 \label{Bs}
\end{align}
and
\begin{align}
 F_+&=A_+\left[ 1-\frac{im_0\cos \phi}{E-\sqrt{E^2-m_0^2}} \right]-A_-\left[  \frac{m_0 \sin \phi}{E+\sqrt{E^2-m_0^2}} \right], \nonumber \\
 F_-&=A_+ \left[ \frac{m_0\sin \phi}{E+\sqrt{E^2-m_0^2}} \right]+A_- \left[ 1-\frac{im_0\cos \phi}{E-\sqrt{E^2-m_0^2}} \right].
 \label{Fs}
\end{align}
With Eqs.~(\ref{Bs}) and~(\ref{Fs}) at hand, we can now write the solutions for $\varPhi_\pm^{1,2}$.
For $z<0$
\begin{align}
 \varPhi_+^{1,2}=A_+ \ e^{iEz}-&\Bigg[ A_+\left( \frac{im_0\cos \phi}{E+\sqrt{E^2-m_0^2}} \right) \nonumber \\
 &+A_-\left( \frac{m_0\sin \phi}{E+\sqrt{E^2-m_0^2}} \right) \Bigg] e^{-iEz}, \nonumber \\
 \varPhi_-^{1,2}=A_- \ e^{iEz}+&\Bigg[ A_+\left( \frac{m_0\sin \phi}{E+\sqrt{E^2-m_0^2}} \right) \nonumber \\
 &-A_-\left( \frac{im_0\cos \phi}{E+\sqrt{E^2-m_0^2}} \right) \Bigg] e^{-iEz}.
 \label{solzneg}
\end{align}
whereas for $z>0$
\begin{align}
 \varPhi_+^{1,2}=&\Bigg [ A_+\left(1- \frac{im_0\cos \phi}{E-\sqrt{E^2-m_0^2}} \right) \nonumber \\
 &-A_-\left( \frac{m_0\sin \phi}{E+\sqrt{E^2-m_0^2}} \right) \Bigg] e^{i\sqrt{E^2-m_0^2}z}, \nonumber \\
 \varPhi_-^{1,2}=&\Bigg[ A_-\left(1- \frac{im_0\cos \phi}{E-\sqrt{E^2-m_0^2}} \right) \nonumber \\
 &+A_+\left( \frac{m_0\sin \phi}{E+\sqrt{E^2-m_0^2}} \right) \Bigg] e^{i\sqrt{E^2-m_0^2}z}.
 \label{solzpos}
\end{align}
We now use these solutions to find the transmission and reflection coefficients.

\section{\label{IV}Transmission and reflection coefficients}
 
To write the solution of the Dirac equation, we substitute Eqs.~(\ref{solzneg}) and~(\ref{solzpos}) into Eq.~(\ref{decomposition}). In order to get the incident $\Psi^{inc}$, reflected $\Psi^{ref}$ and transmitted $\Psi^{tra}$ waves, we first compute Eq.~(\ref{soldiraceq}). We start this procedure with the incident wave. This travels from $z<0$ to $z=0$ and therefore we obtain it from
\begin{eqnarray}
 \Psi^{(inc)}&=&\{ \gamma^0 E-i\gamma^3\frac{\partial}{\partial z}\} \nonumber \\
 &\times& \{ \varPhi_+^1(z)u_+^1+\varPhi_-^1(z)u_-^1+\varPhi_+^2(z)u_+^2+\varPhi_-^2(z)u_-^2 \}, \nonumber \\
 \label{incphi}
\end{eqnarray}
where $\varPhi_\pm^{1,2}$ only includes terms proportional to $e^{iEz}$. Since the reflected wave travels in the opposite direction of the incident one, that is, toward negative values of $z$, we can get it from Eq.~(\ref{incphi}) as well. However, in this case, we take into account only terms in $\varPhi_\pm^{1,2}$ proportional to $e^{-iEz}$. Finally, for the transmitted wave, we consider the solution of $\varPhi_\pm^{1,2}$ for positive values of $z$. Therefore we get it from
\begin{eqnarray}
 \Psi^{(tra)}&=&\{ \gamma^0 E-i\gamma^3\frac{\partial}{\partial z}+m_0\} \nonumber \\
 &\times&\{ \varPhi_+^1(z)u_+^1+\varPhi_-^1(z)u_-^1+\varPhi_+^2(z)u_+^2+\varPhi_-^2(z)u_-^2 \}. \nonumber \\
\end{eqnarray}
 \begin{figure}[t!]
 \begin{center}
 \includegraphics[scale=0.5]{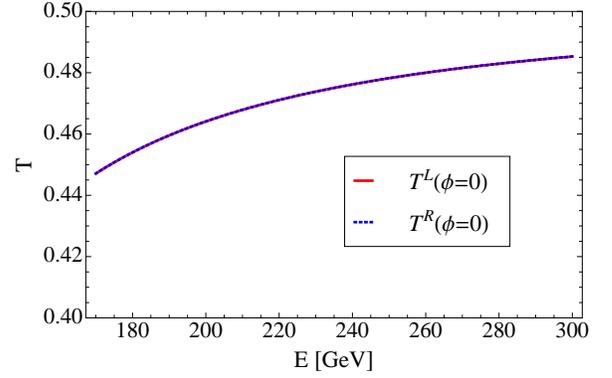}
 \end{center}
 \caption{Transmission coefficients $T^L$ and $T^R$ for $\phi=0$ and $m_0=175$ GeV.}
 \label{fig1}
 \end{figure}
To quantify the probability that particles are either transmitted or reflected, we need to compute the corresponding reflection and transmission coefficients. These are built from the incident, reflected and  transmitted currents. Recall that for a given wave function $\Psi$, the current normal to the wall is given by
\begin{equation}
 J=\Psi^\dagger \gamma^0 \gamma^3 \Psi,
\end{equation}
\noindent
where the incident, transmitted and reflected currents are
\begin{align}
 J_{inc}&=-8(A_+^2+A_-^2)E^2, \nonumber \\ 
 J_{ref}&=8(A_+^2+A_-^2) \nonumber \\ &\times\frac{E^2m_0^2}{2E^2-m_0^2+2E\sqrt{E^2-m_0^2}}, \nonumber \\
 J_{tra}&=-16(A_+^2+A_-^2) \nonumber \\
 &\times\frac{E^2\sqrt{E^2-m_0^2}(E+\sqrt{E^2-m_0^2})}{2E^2-m_0^2+2E\sqrt{E^2-m_0^2}},
 \label{currents}
\end{align}
respectively. To write Eq.~(\ref{currents}), we have kept the currents expressed in terms of the incident amplitudes $A_\pm$. However. since we are only looking for scattering states, \textit{i.e.}, positive energies, $A_-=0$, and the currents become
\begin{align}
 J_{inc}&=-8A_+^2E^2, \nonumber \\ 
 J_{ref}&=8A_+^2\frac{E^2m_0^2}{2E^2-m_0^2+2E\sqrt{E^2-m_0^2}}, \nonumber \\
 J_{tra}&=-16A_+^2\frac{E^2\sqrt{E^2-m_0^2}(E+\sqrt{E^2-m_0^2})}{2E^2-m_0^2+2E\sqrt{E^2-m_0^2}}. 
 \label{explicitcurrents}
\end{align}
Since the total current is conserved, $J_{inc}=J_{ref}+J_{tra}$, we have
\begin{equation}
 1=\frac{J_{ref}}{J_{inc}}+\frac{J_{tra}}{J_{inc}},
\end{equation}
with $R\equiv -J_{ref}/J_{inc}$ (the minus sign accounts for the fact that the incident and reflected currents travel in opposit directions) and $T\equiv J_{tra}/J_{inc}$. From Eq.~(\ref{explicitcurrents}), the reflection and transmission coefficients, $R$ and $T$, become
\begin{align}
 R&=\frac{m_0^2}{2E^2-m_0^2+2E\sqrt{E^2-m_0^2}}, \nonumber \\
 T&=\frac{2\sqrt{E^2-m_0^2}(E+\sqrt{E^2-m_0^2})}{2E^2-m_0^2+2E\sqrt{E^2-m_0^2}}.
 \label{coeff}
\end{align}
 \begin{figure}[t!]
 \begin{center}
 \includegraphics[scale=0.5]{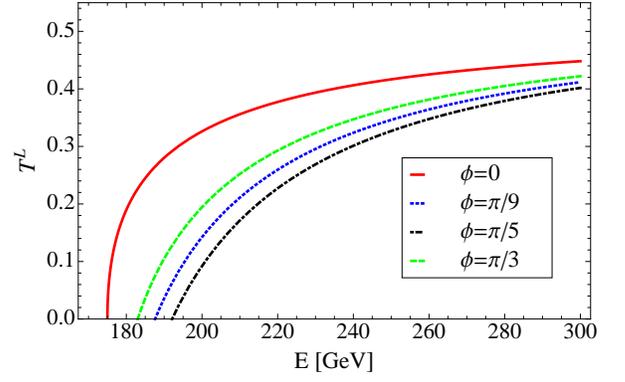}
 \end{center}
 \caption{Transmission coefficient $T^L$ for four different values of $\phi$ and $m_0=175$ GeV.}
 \label{fig2}
 \end{figure}
Recall that for particles in the false vacuum, the equations of motion for left- and right-handed modes are the same since they have vanishing mass. The distinction between modes is made evident in the true vacuum, once they have crossed the wall, where their equations of motion are different. Therefore, in the region $z>0$, it is possible to separate the transmitted current into left- and right-handed, namely,
\begin{align}
J_{tra}^L&=\Bigg [\frac{1}{2}(1-\gamma_5)\Psi_{tra} \Bigg ]^*\gamma^0 \gamma^3 \Bigg [\frac{1}{2}(1-\gamma_5)\Psi_{tra}\Bigg ], \nonumber \\
J_{tra}^R&=\Bigg [\frac{1}{2}(1+\gamma_5)\Psi_{tra} \Bigg ]^*\gamma^0 \gamma^3 \Bigg [\frac{1}{2}(1+\gamma_5)\Psi_{tra} \Bigg ].
\label{transcurrentLR}
\end{align}
The currents are given explicitly by
\begin{eqnarray}
 J_{tra}^L&=&\frac{4A_+^2}{2E^2-m_0^2+2E\sqrt{E^2-m_0^2}} \nonumber \\ 
 &\times& \Big[ -2E^2\Big( E^2-m_0^2+E\sqrt{E^2-m_0^2} \Big) \nonumber \\
 &+&4Em_0^3\cos^2\phi \sin \phi \Big], \nonumber \\
 J_{tra}^R&=&\frac{4A_+^2}{2E^2-m_0^2+2E\sqrt{E^2-m_0^2}} \nonumber \\ 
 &\times& \Big[ -2E^2\Big( E^2-m_0^2+E\sqrt{E^2-m_0^2} \Big) \nonumber \\
 &-&4Em_0^3\cos^2\phi \sin \phi \Big].
 \label{bothcurrents}
\end{eqnarray}
 \begin{figure}[t!]
 \begin{center}
 \includegraphics[scale=0.5]{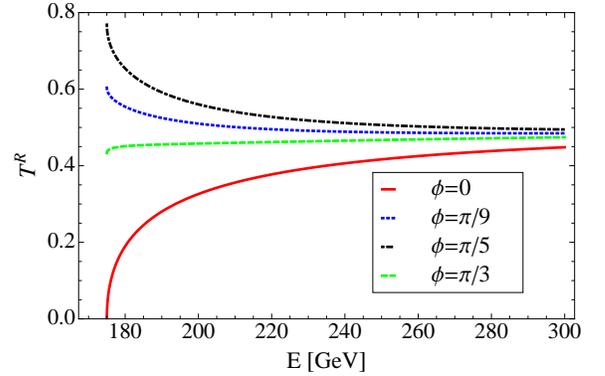}
 \end{center}
 \caption{Transmission coefficient $T^R$ for four different values of $\phi$ and $m_0=175$ GeV.}
 \label{fig3}
 \end{figure}
From Eq.~(\ref{bothcurrents}) it is clear that the transmitted currents for the two modes are different due to the term $\pm 4Em_0^3\cos^2\phi \sin \phi$. We can also check that when $\phi=0$, the transmitted currents become equal. To obtain the corresponding transmission coefficients, we divide $J_{tra}^L$ and $J_{tra}^R$ by the incident current, namelly,
\begin{align}
 T^L&=\frac{E^3-Em_0^2+E^2\sqrt{E^2-m_0^2}-2m_0^3\cos^2\phi \sin \phi}{2E^3-Em_0^2+2E^2\sqrt{E^2-m_0^2}}, \nonumber \\
 T^R&=\frac{E^3-Em_0^2+E^2\sqrt{E^2-m_0^2}+2m_0^3\cos^2\phi \sin \phi}{2E^3-Em_0^2+2E^2\sqrt{E^2-m_0^2}}.
 \label{TLTR}
\end{align}

 \begin{figure}[b!]
 \begin{center}
 \includegraphics[scale=0.5]{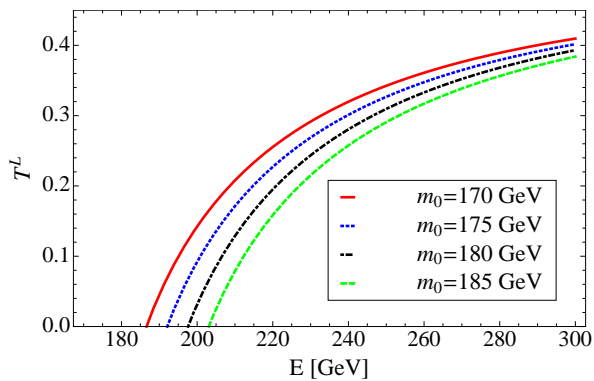}
 \end{center}
 \caption{Transmission coefficient $T^L$ for four different values of $m_0$ evaluated for $\phi=\pi/5$.}
 \label{fig4}
 \end{figure}
 
\begin{figure}[t]
 \begin{center}
 \includegraphics[scale=0.5]{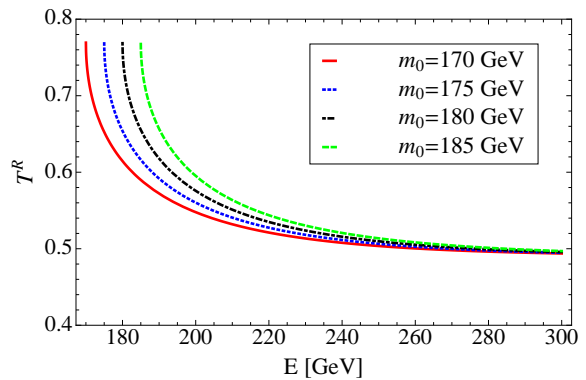}
 \end{center}
 \caption{Transmission coefficient $T^R$ for four different values of $m_0$ evaluated for $\phi=\pi/5$.}
 \label{fig5}
 \end{figure}

Equation~(\ref{TLTR}) contains the information on the fraction of incoming fermions passing through the wall out of the incident ones.  The asymmetry is quantified in terms of the phase $\phi$. Figure~\ref{fig1} shows the transmission and reflection coefficients for the case when $\phi=0$ as a function of the incident particle's energy.  As expected, for this case both $T^{L,R}$ coincide. Figure~\ref{fig2} shows the left-handed transmission coefficient for four different values of $\phi$. As the phase increases from 0 to the critical value $\phi_c\simeq \pi/5$ the curves become flatter for low energies.  For values $\pi/5 < \phi < \pi/2$ the curves become steeper. The behavior of the right-handed transmission coefficient is shown in Fig.~\ref{fig3}. In this case the curvature changes from concave to convex as the phase increases from 0 to the critical value $\phi_c\simeq \pi/5$. For values $\pi/5 < \phi < \pi/2$ the curves become concave again. For $\phi=\phi_c$ the asymmetry is maximum and it vanishes again for $\phi=\pi/2$. Notice also that since the factor responsible for the asymmetry depends on the particle's mass, the transmission and reflection coefficients as functions of the particle's energy are affected by the value of $m_0$; the asymmetry is magnified as $m_0$ increases. This is shown in Fig.~\ref{fig4} (\ref{fig5}) for the case of the left-handed (right-handed) transmission coefficient.

\section{\label{V}Summary and Conclusions}

 \begin{figure}[t!]
 \begin{center}
 \includegraphics[scale=0.5]{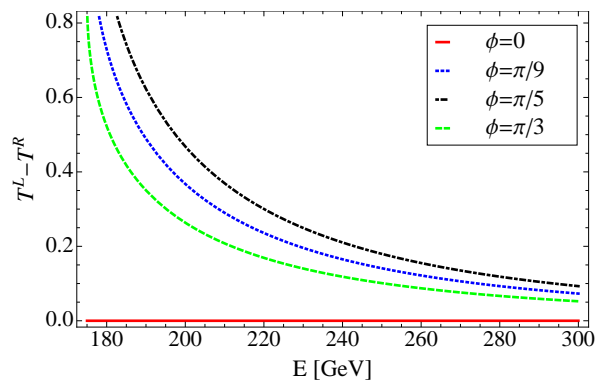}
 \end{center}
 \caption{Strength of net generated electric current in units of the fermion's charge absolute value $|q_f|$ quantified in terms of the difference between the transmission coefficients $T^L - T^R$ for $m_0=175$ GeV and several values of the phase $\phi$. This electric current is locally generated just after fermions pass through the bubble wall.}
 \label{fig6}
 \end{figure}

In conclusion, we have studied in detail the asymmetrical transmission of fermions incident on EWPT bubble walls from the false into the true vacuum. The wall is considered in the infinitely thin limit and modeled as a step  function whose hight is the particle's mass. A CP-violating interaction term between fermions and the wall is included by means of a complex phase factor and quantified by the phase $\phi$. We have shown that for $0<\phi < \pi/2$, an asymmetry in the transmission of left- and right-handed modes is developed. 
 
Since CP is violated, the above implies that for fermions with a charge $q_f$, a net local electric current transverse to the wall
\begin{equation}
 \vec{J}=q_f\left(\vec{J}_{tra}^L-\vec{J}_{tra}^R\right),
\end{equation}
\noindent
is generated in the true vacuum. Figure~\ref{fig6} shows the strength of such current, quantified in terms of the difference between the transmission coefficients $T^L - T^R$ just after fermions pass through the wall. 

Recall that an important property of a magnetic field is its helicity. This is defined as the integral over a closed volume (one where the magnetic field lines are fully contained) of the dot product between the magnetic field $\vec{B}$ and the vector potential $\vec{A}$. With this definition, helicity is a gauge invariant quantity. A homogeneous magnetic field has zero helicity. Notice that the scenario advocated in this work produces electric currents normal to the bubble walls. The intensity of these currents at each scattering point on the bubble wall depends on the energy of the incident particle and thus it is a probabilistic process. When these bubbles are considered as three-dimensional spheres, the produced currents are radial and have different intensities. Therefore the induced magnetic field can be helical. This is an important feature since turbulent diffusion cannot dissipate helicity~\cite{Brandenburg}. Hence, with the kind of magnetic fields produced in this scenario, it becomes possible that during the universe's evolution, part of the magnetic field energy can be transferred from shorter to longer wavelengths (inverse cascade), thus avoiding that the fields become short ranged~\cite{quickly}.

Notice also that the strength of the electric currents, and thus of the magnetic field generated by this mechanism, is proportional both to the (product of the square of the cosine times the sine of the) CP violating phase between left- and right-handed modes as well as to the intensity of the Yukawa interaction between fermions and the Higgs, represented by the value $m_0$, hereby chosen to correspond to the top quark. The asymmetry is maximal for $\phi\simeq \pi/5$. Such a magnitude of a CP-violating phase is comparable to the size of the complex phase in the CKM matrix~\cite{PDG}. Therefore, to study its subsequent evolution, one can think of remaining within the realm of the SM. Nevertheless it is known that the intensity of the interaction is not enough to produce a strong enough first order phase transition to avoid the sphaleron erasure of the baryon asymmetry (BAU) produced during the EWPT~\cite{Cline}. Although this intensity may be enough to generate a strong enough magnetic field seed, to strengthen the interaction, and thus the magnitude of the generated field, and perhaps link magnetogenesis with the BAU during the EWPT, it may be interesting to explore extensions of the SM such as two Higgs doublet models or even models with a singlet scalar coupling to the Higgs.  Work along these lines is in progress and will be reported elsewhere.

\section*{Acknowledments}

Support for this work has been received in part by UNAM-DGAPA-PAPIIT grant number IN101515 and by Consejo Nacional de Ciencia y Tecnolog\1a grant number 256494.

\end{document}